\documentstyle[psfig]{caosp}
\begin{document}
\pubyear{1998}
\volume{27}

\firstpage{338}
\htitle{Observations of roAp stars at the Mt. Dushak-Erekdag...}
\hauthor{T.N. Dorokhova, N.I. Dorokhov }
\title {Observations of roAp stars at the Mt. Dushak-Erekdag
station of Odessa Astronomical Observatory}
\author{T.N. Dorokhova, N.I. Dorokhov}
\institute{ Department of Astronomy, Odessa State University,
Odessa 270014 Ukraine\\}
\maketitle

\begin{abstract}
Since 1992, observations of roAp stars have been carried out using the
dual-channel photometer attached to the 0.8m telescope, which is situated
in Central Asia, at the Mt. Dushak-Erekdag station of Odessa Astronomical
Observatory. Some results of observations of $\gamma$ Equ and
of HD 134214 are presented. 5 stars were investigated as roAp candidates.
The Fourier spectra of 4 stars did not show any variability in the
high-frequency region. The Fourier spectrum of HD 99563 revealed a peak at a 
frequency f=128.9 c/d and with a semi-amplitude of 3.98 mmag.
\end{abstract}

Rapidly oscillating Ap stars were discovered by D. Kurtz in 1978
(Kurtz, 1990).
Up to now the list of roAp stars comprises 28 names, and nearly all of the
stars were discovered in SAAO (see Kurtz 1990, Martinez et al. 1991,
Martinez \& Kurtz 1994a, b). Therefore, most of the objects were detected in
the southern hemisphere. Although there were several surveys aimed at
discovering especially northern roAp stars (see for example
Nelson \& Kreidl 1993), only two stars from the list have positive
declinations.

  We used for roAp stars' observations
the dual-channel photometer mounted on the 0.8m telescope
 at the Mt. Dushak-Erekdag station (Central Asia)
of Odessa Astronomical Observatory (Dorokhov et al., 1995).

 In 1992  Mt. Dushak-Erekdag station participated in the multisite photometric
campaign on $\gamma$ Equ  organised by T. Krejdl and M. Nelson (see Martinez
et al., 1996).
We continued observations of $\gamma$ Equ in the years 1993 and 1994.
It would be interesting to investigate the variability
on short timescales corresponding to modes with extremely low pulsational 
amplitudes.

 In 1993 we  observed HD 134214 simultaneously in
Str\"omgren's v-filter and Johnson's R-filter, using a split-beam prism
on cloudy-windy nights to test the possibility
of dual-channel photometry.
It appeared that atmospheric variations have a common
tendency through both filters at low frequencies, but are different at
frequencies higher than 50 c/d: their amplitudes are lower and the 
characteristic times somewhat longer in the red spectral region than
in the blue one (Dorokhov et al., 1996).

 The search for new roAp stars in the northern hemisphere is
particularly important to balance the number of southern and northern roAp 
stars. Our search was somewhat arbitrary at the beginning, then
we used the photometric criteria by Martinez et al. (1991),
extracted the list of candidates from the uvby$\beta$ Catalogue (Hauck \& 
Mermilliod, 1990) and verified the list with the Catalogue of Ap and
Am stars (Renson, 1991).

The observations were episodic, because the Mt. Dushak-Erekdag
station is located three thousand kilometers away from Odessa, and we visited
the station to participate in international programs or multisite
campaigns. During 4 years we were able to observe 5 stars of the candidates'
list: HD 15257, HD 17317, HD 99563, HD 115606, HD 217401.

The data were acquired as continuous 10 or 20 sec integrations,
a comparison star was observed simultaneously in the second channel,
whenever a suitable star occurred in the field of view of the telescope.
2 - 6 hours' series per star in Johnson's B or Str\"omgren's v filters
were obtained in one or two nights.
The Fourier spectra of 4 stars did not show any variability larger than
the 1 - 1.2 mmag level in the frequency region 50 - 400 c/d.
The Fourier spectrum of HD 99563 revealed a peak at the frequency 127.6 c/d.
Normalized data yielded a more prominent peak
at the frequency f=128.9$\pm0.635$ c/d (semi-amplitude
3.978$\pm0.38$ mmag) (Fig.1).

\unitlength=1in
\begin{figure}
\begin{center}
\psfig{figure=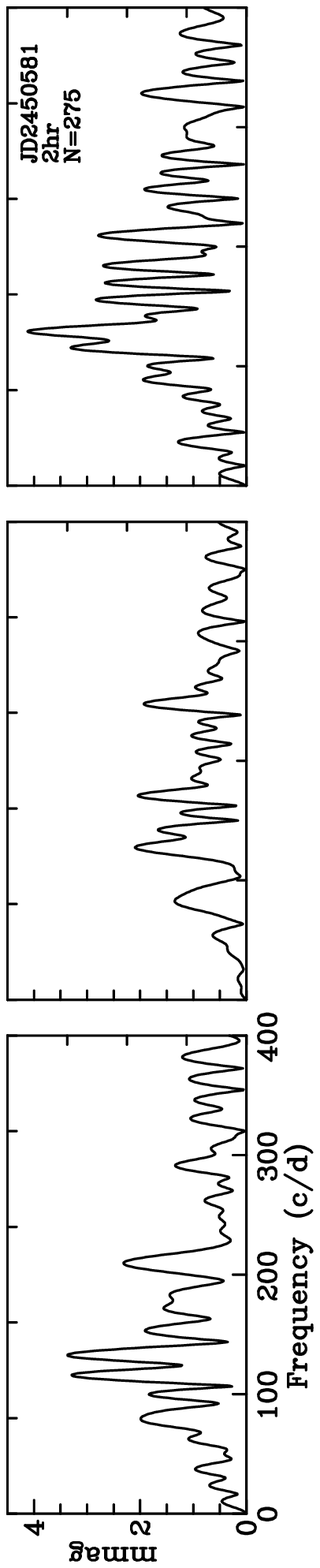,angle=270,height=1.2in}
\psfig{figure=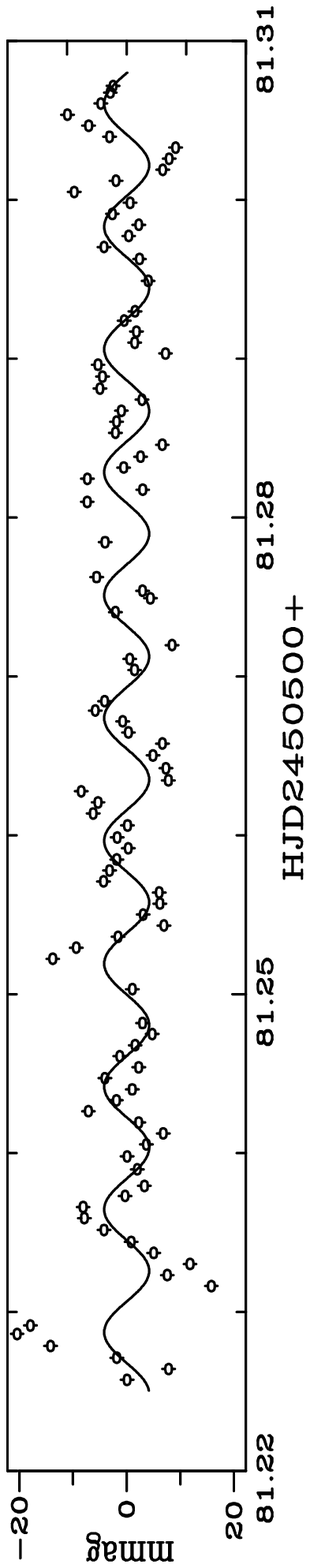,angle=270,height=1.2in}
\caption{{\bf Top:} Fourier spectra of HD 99563 (left), comparison star
(centre) and Fourier spectrum of differences (right). {\bf Bottom:} Light
curve of HD 99563 with the best-fit sinusoid for the 11.2 min period.}
\end{center}
\end{figure}

 The bottom panel in Fig. 1 shows 2 hours' lightcurve of the star.
 The results of the work may be considered as only preliminary.
We share the cautions made by Martinez \& Kurtz (1994b), that roAp stars
must be observed in the best atmospheric conditions and the observations
should be continued until reliable results are achieved repeatedly.

\end{document}